\documentclass[showpacs,preprintnumbers,amsmath,
amssymb]{revtex4}
\usepackage{graphicx}
\usepackage{dcolumn}
\usepackage{bm}
\usepackage{tensor}
\begin{document} 

\title{Twisted superfluids in moving frame.}
\author{A.Yu.Okulov}
\email{alexey.okulov@gmail.com}
\homepage{https://sites.google.com/site/okulovalexey}
\affiliation{Russian Academy of Sciences, 119991,  
Moscow, Russian Federation.}
 
\date{\today}

\begin{abstract}
{Helical waveguide filled by superfluid is shown to transform 
rotations of the reference frame $\vec \Omega_{\oplus}$ into 
linear displacements of the atomic ensemble and vise versa the linear 
displacements of the reference frame $\vec V$ initialize rotations 
$\vec \Omega_{\oplus}$ of ensemble. In the 
mean-field Gross-Pitaevskii equation with a weakly 
modulated gravitation potential the exact solutions 
for macroscopic wavefunction $\Psi$ demonstrate  
the emergence of $redshift$  phase factor. For helical waveguides formed 
by phase-conjugated optical vortices with high angular 
momentum $L_z=\pm \ell \hbar \sim \pm \hbar \cdot 10^{3 \div 4}$ one may expect the significant improvement of the 
phase-sensitivity $\delta \phi$ by factor containing $2 \ell$ 
compared to the conventional nonsingular 
matter-wave interferometry.}

\end{abstract} 

\maketitle
 

\vspace{1cm}
 
The conservation laws exist due to the space-time symmetries. 
Homogeneity of time, isotropy and homogeneity 
of space lead to invariance of energy, linear momentum and 
angular momentum of isolated system with respect to infinitesimal translations 
$dt, dz$ and rotations $ d\theta$ \cite {Landau:1982}:
\begin{equation}
\label {symmetry_operators} 
\hat H= i\hbar \frac {\partial}{\partial t},
\hat P_z= -i\hbar \frac {\partial}{\partial z},
\hat L_z= -i\hbar \frac {\partial}{\partial \theta},{\:}
\end{equation} 
where $(t,z,r,\theta)$ is cylindrical coordinate system, 
$\hat H,\hat P_z,\hat L_z$ are operators of energy, 
linear and angular momenta projections on motion axis $z$.
Lorenz invariance states that observer in closed box placed 
in inertial reference 
frame is not able to detect his motion with respect to 
outer world unless acceleration to appear  
\cite {Michelson:1904}. 
In the same way the isotropy of space gives no chance to find  
preferential direction with a special properties. 
For these reasons isolated frictionless setup in isotropic 
space demonstrates 
mutual exchange of angular momentum between its constituents 
provided the external torque is absent (fig.1). But when 
the axis of rotating top is deflected forcedly the rest of 
setup will undergo the torque $\vec T$ due to angular momentum 
$\vec L$ conservation. 

	The conservation of angular momentum $\vec L$ means that 
when a part of isolated system in $apriory$ isotropic space 
acquires rotation 
hence it's angular momentum (AM) is changed the rest of system should acquire
the oppositely directed AM. The most popular example is 
experimentalist on rotating platform with rotating $top$ in his 
hands (fig.1a) \cite{Feynman:1963}. When 
rotating $top$ with $\vec L_{top}(t)$  is tilted 
with respect 
to rotation axis $\vec L_{plat}(t)$ 
the torque $\vec T(t)= - d \vec L_{top}(t) / d t$ appears acting on 
a platform in order to conserve the total angular momentum 
of the system 
$\vec L(t)=\vec L(t=0)=\vec L_{plat}(t) + \vec L_{top}(t)$. 
When at $t=T$ $top$ goes to exactly opposite direction at the angle $180^o$  
the platform acquires doubled angular momentum 
$\vec L_{top}(t=0)= -\vec L_{top}(T)$
of the top $\vec L_{plat}(t=T)= 2 \vec L_{top}(t=0)$.

	Another example is reversal of 
photon's spin $S_z=\pm \hbar$ in Beth's experiment (fig.1b) 
\cite{Beth:1936}. The visible light with  
circular polarization propagates twice through $\lambda/2$ 
plate suspended on quartz wire. As a consequence  
torque $\vec T_{\lambda/2}$ on a plate appears whose 
value $|\vec T_{\lambda/2}|=2 \it {I} S/\omega$ 
is proportional to light intensity $\it {I}$, 
where $S$ is the square of plate, $\omega$ is 
carrier frequency of light. Birefringent $\lambda/4$ plate 
knocks over the $\pm \hbar$ angular momentum of photon 
to the opposite one $\mp \hbar$ in double pass after 
reflection from mirror $M$.
The important feature of 
Beth's experiment is photon's $spin$ $reversal$ 
in reflection from mirror $\bf M$: for this purpose 
the $\lambda/4$ plate is placed near $\bf M$ in order 
to alternate $spin$ $\mp \hbar$ of reflected photons. Without 
$\lambda/4$ plate the optical  
torque $\vec T_{\lambda/2}$ would be exactly zero \cite{Beth:1936}. 
The $\lambda/4$ plate and mirror $\bf M$ double 
the torque upon 
$\lambda/2$ plate due to $phase$-$conjugation$ of photons 
\cite {Basov:1980}.

	The else example is from the field of singular optics 
\cite {Volostnikov:1989} where structured 
light beams with orbital angular momentum $L_z=\pm \ell \hbar$ (OAM) and 
topological charge $\ell$ 
\cite {Allen:1992} interact with anisotropic media. 
Such interaction induces torque on medium and exchange of angular momenta
between photons and medium. 
The unique example of $perfect$ OAM $reversal$ is optical phase conjugation 
\cite {Okulov:2008}(fig.1c). Phase conjugation means a perfect 
coincidence of helical phase 
fronts of incident and reflected photons hence angular momentum 
of phase-conjugated photons $L_z=\mp \ell \hbar$ 
ought to be exactly opposite to OAM of 
the incident ones $L_z=\pm \ell \hbar$. As a result of reflection 
the anisotropic phase-conjugating mirror (PCM) acquires doubled angular momentum $L_z=\pm 2 \ell \hbar$ 
as a recoil.

	Current communication aims to show that 
small perturbations of 
the helical waveguide orientation and displacements 
\cite{Okulov:2013} are capable to produce   
macroscopically observable motions due to exchange of angular 
momenta between superfluid. 
Indeed superfluid localized 
within twisted boundaries where rotational symmetry is absent 
turns to be sensitive to displacements 
with translational speed $\vec V$, 
rotations $\vec \Omega_{\oplus}$  and acceleration $\vec g$ 
of reference frame where waveguide is in rest(fig.1d). 
This situation may be realized experimentally by virtue of  
quantum interference \cite{Malomed:2020} device with 
twisted interference fringes \cite{Okulov:2012}. 
Atomic interference fringes 
are formed inside helical waveguide produced by two phase-conjugated 
optical vortices with carrier frequencies detuned by $\delta \omega$ 
\cite{Okulov:2008J}. Thus twisted waveguide potential $U_{tw}$ is given 
in cylindrical coordinates  $t,z,r, \theta$ as follows \cite{Lembessis:2018}:

\begin{equation}
\label{potential_rot_frame}
{ U_{tw}{(t,z,r,\theta)}} \sim 
\exp \bigg{[}{\frac {-2r^2}{D(z)^2}}\bigg{]}
{[1+ \cos (2kz+2\ell\theta+\delta \omega t )]}\cdot
{\frac{r^{2|\ell|}}{D(z)^{2|\ell|}}} , {\:}
\end{equation}
where Rayleigh range $L_R \cong  D_0^2/{\lambda}$ of 
$LG$ (Laguerre-Gaussian) vortex beams 
with topological charge $\ell$ \cite{Fiona:2014} 
depends on beam waist radius $D_0$,  wavenumber $2 \pi /\lambda$,
$D(z)=D_0 \sqrt{1+{ {z^2}/{k^2 D_0^4}}}$.
Such helical trapping potential had been already implemented experimentally 
with phase conjugating photorefractive mirror 
\cite{Saffman:1996,Woerdemann:2009} 
and phase-conjugating mirror using four-wave mixing in trapped 
atomic ensemble \cite{Tabosa:1999}.Noteworthy this OAM $reversal$ 
holds for speckle fields \cite {Okulov:2009} as well. 
In a given speckle pattern OAM carried by multiple 
vortices is inverted $locally$ for each vortex in the speckle provided 
each inversion of $L_z=\pm \ell \hbar$ results in excitation 
of acoustical vortex with $L_z=\pm 2\ell \hbar$ \cite {Okulov:2010plasma}. 

	Consider the general case when matter wave interferometer is
placed upon rigid gravitating 
sphere rotating with angular velocity $\vec \Omega_{\oplus}$ and 
this sphere moves also translationally with linear velocity $\vec V$.
(fig.2). 
The interferometer 
axis $\vec z (t)$ (unit dimensionless vector) moves around  
frame rotation axis $\vec \Omega_{\oplus}$ 
and it is in translational motion 
with speed $\vec V$ simultaneously. Two reference frames 
will be considered hereafter. 
In a reference frame collocated 
with interferometer ($lab$ $frame$) the Gross-Pitaevskii 
equation \cite{Pitaevskii:1999} 
for atomic ensemble wavefunction $\Psi$ has the form :
\begin{eqnarray}
\label{GPE_rot_frame}
{i \hbar}{\frac {\partial {\Psi}}
{\partial t}} = -{\frac {\hbar^2}{2 m}} 
\Delta {\Psi} +{ \tilde U_{tw}{(z,r,\theta)}}{\:}{\Psi} 
+{\tilde g}N_a| {\Psi}|{\:}^2 {\Psi}
&& \nonumber \\
-\vec V \cdot \vec z {\:}\hat {P_z} {\Psi}
-\vec \Omega_{\oplus} \cdot \vec z {\:}\hat{L_z} {\Psi}+
m z {\:}
{[\vec g_0 + \epsilon \vec g_1(t)]}\cdot {\vec z}{\:}{\Psi}, {\:}
\end{eqnarray}
where $m$ is the 
mass of trapped atom, $N_a$ is number of atoms in condensate, 
${\tilde g}=4 \pi \hbar^2{\:}a_S/m$, $a_S$ is S-wave scattering 
length,  
$\vec g=\vec g_0+\epsilon \vec g_1(t)$ is free-fall acceleration 
slowly perturbed in time $t$, $\epsilon$ is small parameter,
$\mu$ is chemical potential 
\cite{Lembessis:2016}. 
In Thomas-Fermi approximation the exact solution 
of this equation  
is a superposition 
of the two phase-conjugated matter-wave vortices:
\begin{eqnarray} 
\label{matter waves PC}
{ \tilde \Psi_{tw}{(t,z,r,\theta )}} \sim 
\exp {\bigg[\frac{-i \mu t + i m z \int^t_{-\infty} \vec g(\tau) 
\cdot {\vec z} d \tau }{\hbar} \bigg ]}\cdot 
&& \nonumber \\
{\bigg[{ \cos (kz+\ell\theta+ 2 k t {\:}\vec V \cdot \vec z 
+ 2 {\:} \ell t {\:}\vec \Omega_{\oplus} \cdot \vec z  ) }\bigg{]}}
{\frac{r^{|\ell|}}{D(z)^{|\ell|}}}. {\:}
\end{eqnarray}

This solution for macroscopic 
wavefunction in helical potential predicts that 
from the point of view of $external$ $observer$ , i.e. when viewed from a so-called 
$rest$ $frame$, 
spiral atomic cloud with density profile 
$|{\tilde \Psi_{tw}}|{\:}^2=\mu-\tilde U_{tw}{(z,r,\theta)}
{\:} /{\tilde g}$ 
will move along interferometer axis $\vec z$ 
with translational velocity $\vec V \cdot \vec z $  and 
it will rotate with angular velocity 
$\vec \Omega_{\oplus}\cdot \vec z $  under 
following restriction \cite{Okulov:2013}:  
\begin{equation}
\label {self_similariry matter waves} 
{\:}{\:}{\:}{\:}\vec V \cdot \vec z= -
\vec \Omega_{\oplus} \cdot \vec z {\:}\ell/ k. 
\end{equation}
	Besides from the point of view of 
$external$ $observer$ the wavefunction $\Psi_{tw}$ 
contains the phase factor due to potential 
energy of condensate in a slowly perturbed gravity $\vec g(t)$: 
\begin{eqnarray} 
\label{matter waves to external observer}
{  \Psi_{tw}{(t,z,r,\theta )}} \sim 
\exp \bigg[\frac{-i \mu t + i m  z \int^t_{-\infty} \vec g(\tau) 
\cdot {\vec z} d \tau }{\hbar} \bigg ]\cdot
&& \nonumber \\
{\bigg[e^{ikz+i\ell \theta}+e^{-ikz-i\ell \theta}\bigg]}\cdot
\exp \bigg{[}{\frac {-r^2}{D(z)^2}}\bigg{]}
{\frac{r^{|\ell|}}{D(z)^{|\ell|}}}, {\:}
\end{eqnarray}
 
where phase factor is analogous to $redshift$  which is by definition 
the change of photons frequency in gravitational field.   
In accordance with equivalence principle \cite {Landau:1975} 
acceleration of reference frame $\vec g(\tau)$ is 
equivalent to gravity. 
In our case potential energy of superfluid ensemble in GPE 
(\ref {GPE_rot_frame}) changes the 
phase of wavefunction $\Psi_{tw}$ in 
(\ref {matter waves to external observer}) as well. 
Time-dependent phase of superfluid 
ensemble wavefunction is a key quantity in matter-wave interferometry.  
 
	Two other exact solutions of GPE in moving frame demonstrate 
as frame rotations with $\vec \Omega_{\oplus}$ and frame translations 
with $\vec V$ affect the wavefunction $\Psi$. Indeed the sole translation 
of atomic interferometer with constant speed $\vec V$ 
(without rotation and in the absence of gravity $\vec g$) 
under applicability of GPE: 

\begin{equation}
\label{GPE_trans_frame)only}
{i \hbar}{\frac {\partial {\Psi}}
{\partial t}} = -{\frac {\hbar^2}{2 m}} 
\Delta {\Psi} +{ \tilde U_{tw}{(z,r,\theta)}}{\:}{\Psi} 
+{\tilde g}N_a|{\Psi}|{\:}^2 {\Psi}
-\vec V \cdot \vec z {\:}\hat {P_z} {\Psi}, {\:}
\end{equation}

leads to splitting of eigenfrequencies (linear Doppler ???) 
and rotation of matter wavetrain around $\vec z$ axis (fig.2b): 

 \begin{equation}
\label{matter waves under just translation}
{ \Psi_{tw}{(t,z,r,\theta )}} \sim 
\exp \bigg{[\frac{-i \mu t }{\hbar} \bigg ]}
{ \cos (kz+\ell\theta+ 2 k t {\:}\vec V \cdot \vec z  )}\cdot
\exp \bigg{[ -{\frac {r^2}{D(z)^2}}\bigg]}
{\frac{r^{|\ell|}}{D(z)^{|\ell|}}}. {\:}
\end{equation}

	In the same way the sole rotation 
of atomic interferometer with constant angular velocity 
$\vec \Omega_{\oplus}$ 
(without translation and in the absence of gravity $\vec g$): 

\begin{equation}
\label{GPE_rot_frame_only}
{i \hbar}{\frac {\partial {\Psi}}
{\partial t}} = -{\frac {\hbar^2}{2 m}} 
\Delta {\Psi} +{ \tilde U_{tw}{(z,r,\theta)}}{\:}{\Psi} 
+{\tilde g}N_a|{\Psi}|{\:}^2 {\Psi}
-\vec \Omega_{\oplus} \cdot \vec z {\:}\hat{L_z} {\Psi}, {\:}
\end{equation}

leads to splitting of eigenfrequencies and 
translation of matter wavetrain with speed 
$V_{\Omega_{\oplus}}= \ell {\:}\vec \Omega_{\oplus} \cdot \vec z /k$ 
along $\vec z$ axis (fig.2a): 

\begin{equation}
\label{matter waves under just rotation}
{ \Psi_{tw}{(t,z,r,\theta )}} \sim 
\exp \bigg{[\frac{-i \mu t }{\hbar} \bigg ]}
{ \cos (kz+\ell\theta + 
2 {\:} \ell t {\:}\vec \Omega_{\oplus} \cdot \vec z)
}\cdot
\exp \bigg{[ -{\frac {r^2}{D(z)^2}}\bigg]}
{\frac{r^{|\ell|}}{D(z)^{|\ell|}}}. {\:}{\:}{\:}{\:}
\end{equation}

	In this case the matter wave vortices inside 
helical waveguide 
experience the splitting of energy $\delta E$ due to 
Coriolis effect: 
\begin{equation}
\label {Coriolis shift} 
\delta E (t) = - \hbar \ell \vec {z}(t) \cdot \vec \Omega_{\oplus}, 
{\:}{\:} \alpha(t) = \int_{0}^t d \tau {\:} \delta E (t) /{\hbar},
\end{equation}  
where geometric phase 
$\alpha (t)$ had been acquired via 
transport of rotating matter wave vortices trapped by helical 
waveguide (\ref {potential_rot_frame}) 
along a closed trajectory \cite{Okulov:2023,Berry:1984}.
When axes $\vec z$ and $\vec \Omega_{\oplus}$ are parallel 
to each other the differentiation of 
geometric phase gives the frequency 
shift via angular Doppler effect \cite{Okulov:2012josa,Fiona:2014}.  
For the $lab$ upon surface of rotating hard sphere, this 
happens when $lab$ is placed e.g. on Poles vertically, at equator 
horizontally directed to poles and in a some other geometrically evident 
locations. The orthogonal mutual orientation of $\vec z$ 
and $\vec \Omega_{\oplus}$ means that $lab$ is placed horizontally on 
the Poles, vertically deployed on equator and some other obvious cases. 
The general case of arbitrary angle between translational and rotational 
velocities $\vec z$ and $\vec \Omega_{\oplus}$ is a quantum analog to 
classical $wheel$ $wobbling$ motion \cite{Feynman:1963}. The exact solutions 
(\ref {matter waves to external observer}),
(\ref {matter waves under just translation}),
(\ref {matter waves under just rotation}) are each composed of two
phase-conjugating matter-wave vortices and one may expect that 
their perfectly spiral 
matter-wave patterns will move without distortions provided the motions 
of reference frame are sufficiently small. Though more sophisticated 
analysis of stability \cite{Kevrekidis:2020} of these exact solutions in the 
$wheel$ $wobbling$ regime in the spirit of 
Bespalov-Talanov approach would be of high interest \cite{Okulov:2020}. 
 
	Another example is spatiotemporal evolution of the non-equilibrium
exciton-polariton condensate.
For this ensemble the open dissipative
Gross-Pitaevskii (dGP) equation \cite{Dowling:2016} has the form: 
  
\begin{eqnarray}
\label{open dissipative GPE_rot_frame}
{i \hbar}{\frac {\partial {\Psi_{ep}}}
{\partial t}} = -{\frac {\hbar^2}{2 m_p}} 
\Delta {\Psi_{ep}} +{ U_{tw}{(\vec r)}}{\:}{\Psi_{ep}} 
+{g_{pp}}N_{ep}|{\Psi_{ep}}|{\:}^2 {\Psi_{ep}}
&& \nonumber \\
-\vec \Omega_{\oplus} \cdot \vec z {\:}\hat{L_z} {\Psi_{ep}}+
m_{p} z{\:} 
{[\vec g_0 + \epsilon \vec g_1(t)]}\cdot {\vec z}{\:}{\Psi_{ep}} 
&& \nonumber \\
{\:}-\vec V \cdot \vec z {\:}\hat {P_z} {\Psi_{ep}}
+\frac {i}{2}(P(\vec r)-\gamma - \eta |{\Psi_{ep}}|{\:}^2 )
{\Psi_{ep}}{\:},{\:} 
\end{eqnarray}

where $\Psi_{ep}$ is the order parameter for exciton-polariton superfluid, 
$g_{pp}$ is polariton-polariton interaction constant, 
$m_p$ is polariton mass, $P(\vec r)$ is pumpung rate, 
$\gamma$ is polariton decay rate, $\eta $ is gain saturation. 
Again the Thomas-Fermi solution for the condensate density near equilibrium 
$|{\Psi_{ep}}|{\:}^2=(P(\vec r)-\gamma)/{\eta} $ 
for observer in a reference 
frame moving with angular velocity $\vec \Omega_{\oplus}$, 
translational speed $\vec V $ and acceleration 
$\vec g_0 + \epsilon \vec g_1(t)$ is superposition of the two 
phase-conjugated vortices of exciton-polariton superfluid:
 
\begin{eqnarray} 
\label{exact exciton-polariton superfluid}
{ \tilde \Psi_{ep}{(t,z,r,\theta )}} \sim 
\exp {\bigg[\frac{-i \mu t + i m_{p}  z \int^t_{-\infty} \vec g(\tau) 
\cdot {\vec z}{\:} d \tau )}{\hbar} \bigg ]}
&& \nonumber \\
{\bigg [ \cos(kz+\ell\theta) \bigg ]}
\exp \bigg{[ -{\frac {r^2}{D(z)^2}}\bigg]}
{\frac{r^{|\ell|}}{D(z)^{|\ell|}}}, {\:}{\:}{\:}{\:}{\:}{\:}
\end{eqnarray}

with identical dynamics to (fig.2 ) under external disturbances 
of reference frame and time-modulated $redshift$  of superfluid ensemble. 

The coarse azymuthal interference pattern forming macroscopic wavefunctions $\Psi$ 
(\ref{matter waves PC}), (\ref{exact exciton-polariton superfluid}) provides the 
additional opportunity to enhance sensitivity of matter-wave interferometers.
One may suggest that a higher density of the interference 
fringes ($2 \ell$ per $\lambda/2$) in the helical 
interference pattern \cite {Zeilinger:2015} may enhance 
phase sensitivity of this spiral matter-wave interferometer 
by a factor $(2\ell) ^ \alpha$, where $1 \le \alpha \le 2$: 

\begin{equation} 
\label{structured_coherent_state_limit}
{\:}{\:}{\:} 
(2\ell ) ^ \alpha \delta N \cdot \delta \phi \sim 1; 
 \delta \phi = \frac {1}{(2 \ell) ^ {\alpha}\delta N} =  
\frac {1}{(2 \ell) ^ {\alpha}\sqrt N}. 
\end{equation}

This conjecture may be supported by the following arguments. 
Indeed the uncertainty relation for the number of particles in condensate $N$ and 
its phase $\phi$ puts the fundamental limit on achievable accuracy of the phase 
measurement $\delta \phi$ (\ref {structured_coherent_state_limit}). This limit occurs due 
to noncommutativity of number $\hat N$ and phase $\hat \phi$ 
operators $[\hat N,\hat \phi]=i$ \cite{Barnett_Pegg:1990} that leads to evaluation of the fundamentally 
achievable phase accuracy as $ \delta \phi \sim {1/\sqrt(N)}$. The further possible improvement of accuracy  
is known due to the usage of nonclassical quantum states \cite{Caves:1981,Dowling:1998,Dowling:2008} 
$ \delta \phi \sim {1/N}$. 

The possibility of the more sensitive measurements emerges from $commutativity$ of the number operator $\hat N$ and 
angular momentum operators $\hat L$ and $\hat L_z$. Because $[\hat N,\hat L]=0$ and $[\hat N,\hat L_z]=0$, there exists a 
possibility of statistically $independent$ measurements of the phase $\phi$ and rotations angle $\theta$. 
For this reason the $simultaneous$ measurement of the phase $\phi$ and rotations angle $\theta$ may be organized without 
additional noise induced by measurements. The accurate measurements of the rotation angle $\hat \Theta$ with 
deviation $\delta \theta$ will induce 
uncertainty of angular momentum $\hat L_z=-i \hbar \partial / \partial \theta$ due to 
noncommutativity ($[\hat L_z,\hat \Theta]=-i \hbar$) \cite{Judge:1963}. The fluctuations of $\hat L_z$ because of precise 
measurements of $\theta$ will obey standard uncertainty 
relation $\delta L_z \cdot \delta \theta \ge \hbar/2$ \cite{Barnett:2004}. But noise produced by measurements of 
angle $\theta$ will not result in additional noise in phase $\hat \phi$ measurements because of 
commutativity $[\hat N,\hat L]=0$. 
This fundamental property of quantum measurements of wavefunction phase $\phi=arg \Psi$ and azymuthal rotation 
angle $\theta$ offers the unique opportunity to the further improvement of the standard quantum 
uncertainty of phase measurement \cite{Scully:1997} provided a special implementation of the $\theta$ measurement 
procedure. Namely the $simultaneous$ measurement measurement of the $\phi$  and $\theta$ may reach the 
drastically new level $\delta \phi \sim {2 \ell^{-\alpha}/\sqrt N}$. 

When reference frame rotates with container at angular velocity 
$\Omega_{\oplus}$ the circulation acquires the additional Sagnac 
term:
\begin{equation}
\label{circulation_in_rotframe}
\oint _{\Gamma} \vec v \cdot d \vec r = 
\oint _{\Gamma} {(\vec v - \Omega_{\oplus} \times \vec r )}\cdot d \vec r  
= n \kappa - 2 \Omega_{\oplus} \cdot \vec S,{\:} 
\end{equation}
where $\vec S$ is area enclosed by contour $\Gamma$. Hence 
the phase of wavefunction is shifted by 
$\delta \phi_{SF}=(m / \hbar)2 \Omega_{\oplus} \cdot \vec S$. 
Taking into account de Broglie wavelength of massive particle moving 
with speed $|\vec v|$: $\lambda_B=2 \pi \hbar / (m |\vec v|)$ we get 
the evident similarity of Sagnac shifts for massless photons 
$\delta \phi_{phot}$ and matter 
waves along with their numerical comparison in favour of matter waves: 
\begin{equation}
\label{Sagnac_comparison}
\delta \phi_{SF}=
{\frac {4 \pi \Omega_{\oplus} \cdot \vec S}{\lambda_B |\vec v|}},{\:} 
\delta \phi_{phot}=
{\frac {4 \pi \Omega_{\oplus} \cdot \vec S_{phot}}{\lambda c}}, 
{\frac {\delta \phi_{SF}}{\delta \phi_{phot}}}=
{\frac {\lambda m c {|\vec S|}}{h {|\vec S_{phot}|}}}
= {\frac {m c^2 {|\vec S|}}{h\nu {|\vec S_{phot}|}} }
\sim 10^{10} {\frac { {|\vec S|}}{ {|\vec S_{phot}|}} }, 
\end{equation}
because of smallness of photon's energy 
$h \nu=h \lambda/c \sim 1 eV$ compared 
to the rest mass of a typical atom. This advantage 
of the matter waves is a somewhat diminished by a substantially 
larger area $|\vec S_{phot}|$ enclosed by optical fibers compared 
to atomic waveguides $|\vec S|$  \cite {Scully:1997}. 

	In summary the matter wave interferometer with helical waveguide (fig.2)
\cite{Okulov:2012,Lembessis:2017} 
had been considered in perturbed reference frame. The weak space-time metric modulations 
were taken into account via slightly modulated free-fall 
acceleration ${\vec g(\tau)=\vec g_0 + \epsilon \vec g_1(t)}$. 
The exact solutions 
of the Gross-Pitaevskii equation demonstrate the appearance of 
macroscopically observable motion 
of helically confined superfluids in anisotropic environment when $lab$ frame 
is under externally driven displacements, rotations \cite{Cuevas:2007} and 
accelerations $\vec g(t)$. 
The 
time varying free fall acceleration $\vec g(t)$ induces 
phase-modulation of condensate wavefunction being perfect analog 
of gravitational $redshift$ of photons. 

The possible applications 
are inertial navigation, detection of seismic vibrations and gravitational 
waves \cite{LIGO:2016,Kasevich:2008}. The sensitivity of spiral 
superfluid interferometer 
will be evaluated taking into account quantum 
fluctuations of the particles number $\delta N$ and phase 
of wavefunction $\phi$ \cite{Scully:1997} in another work.

\newpage

\begin{figure} 
\center{ \includegraphics[width=10.5cm]{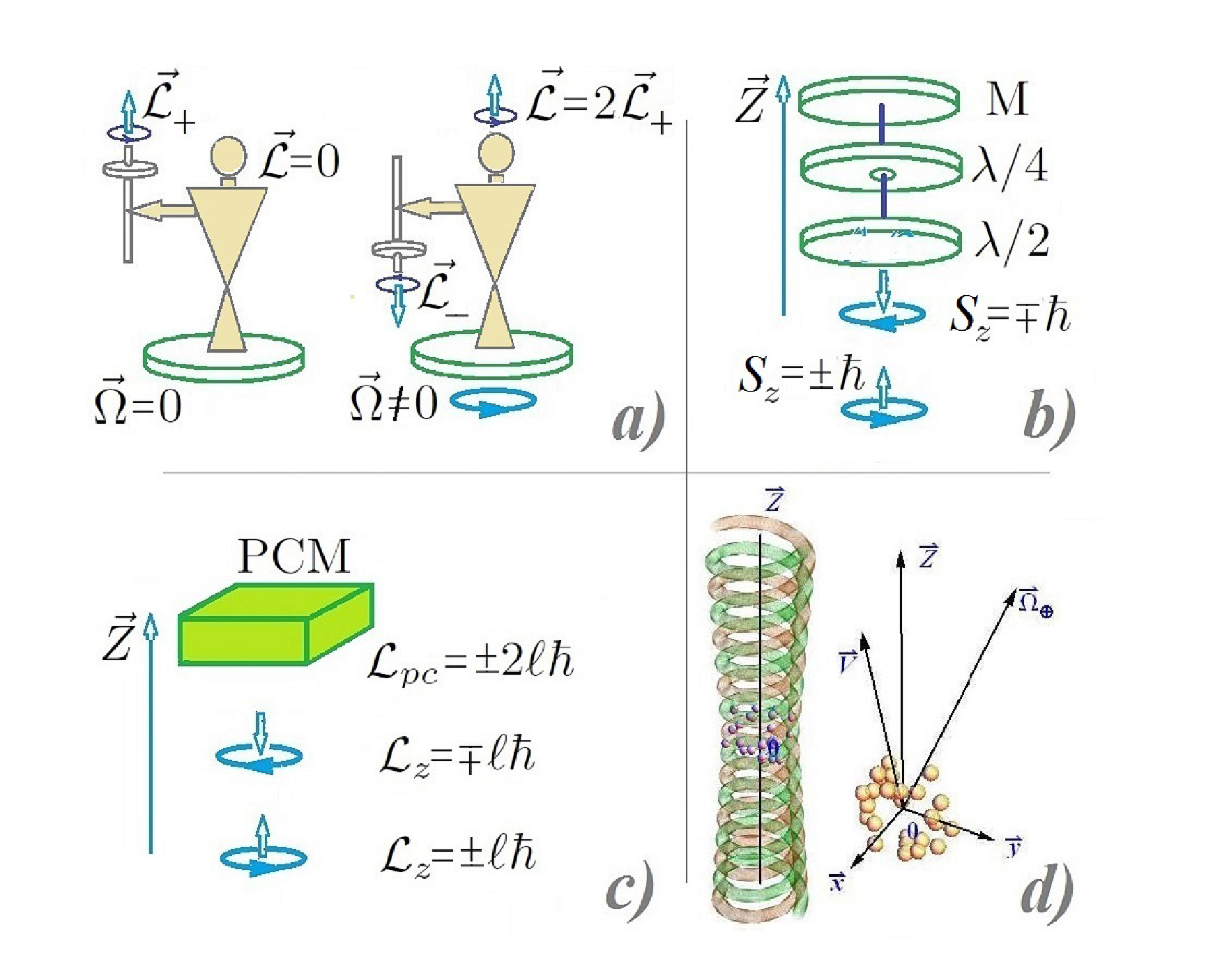}}
\caption{ (Color online) a) Operator upon rotating platform 
reverses the $top$ with 
angular momentum $\vec {\mathcal L}_+$ 
to opposite direction $\vec {\mathcal L}_-$. As a result the platform  
with operator acquire angular momentum $2\vec {\mathcal L}_+$. 
b) $\lambda/2$ quartz plate acquires angular momentum 
$\pm 2\hbar$ due to reversal of photon's angular momentum $\pm \hbar$ in 
Beth's setup. c) Phase-conjugating mirror PCM reverses 
angular momentum of twisted photon $\pm \hbar \ell$ and acquires 
angular momentum $\pm 2\hbar \ell$. 
d) Matter wave interferometer in a frame slightly shaken 
by translations $\vec V$, rotations $\vec \Omega_{\oplus}$ and 
gravity modulations $\vec g(t)$ induces macroscopic motion 
of atomic cloud via "wheel wobbling" interaction.} 
\label{fig.1}
\end{figure}

\begin{figure} 
\center{ \includegraphics[width=7.5 cm]{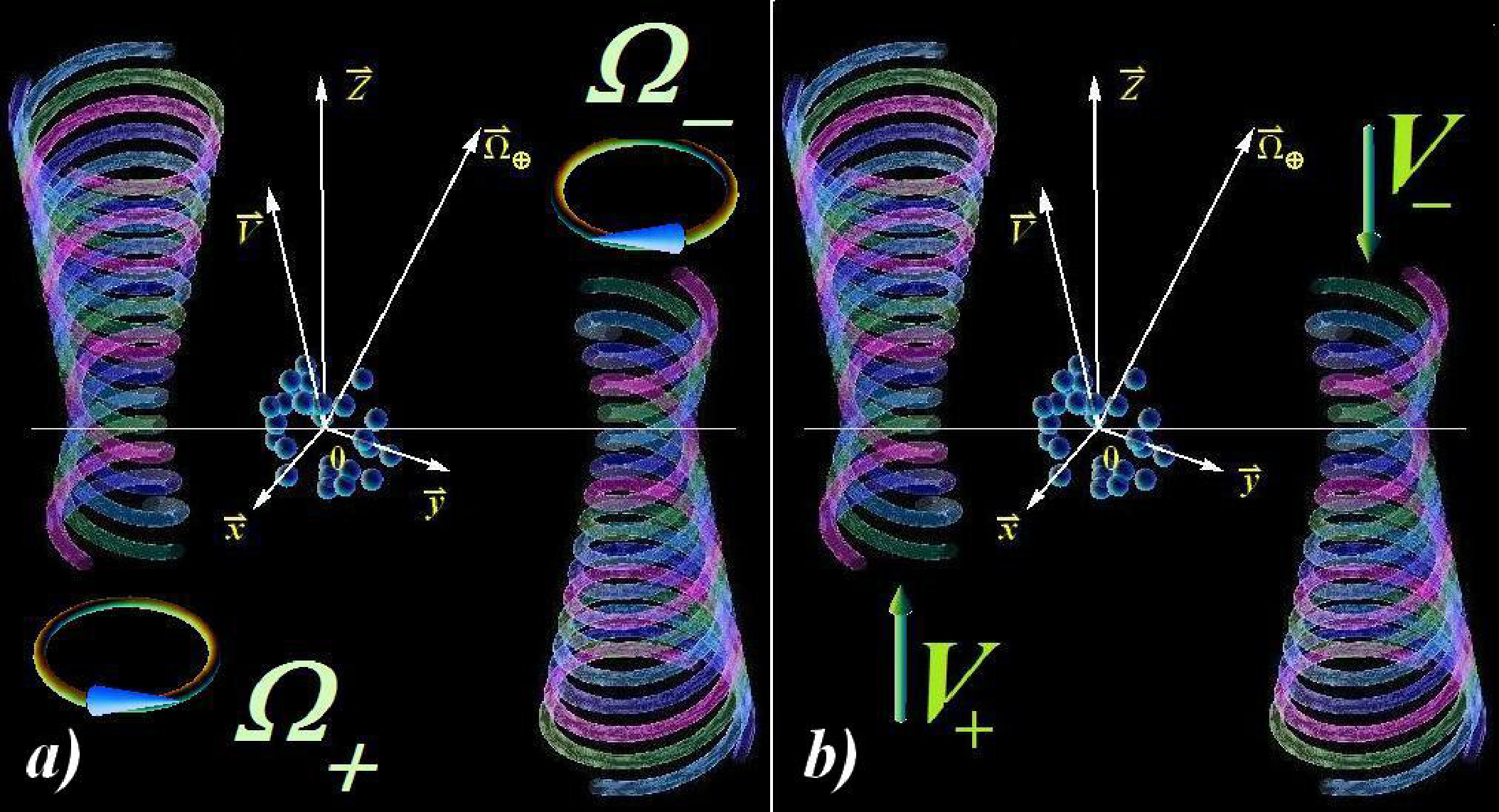}}
\caption{ (Color online) Phase-conjugating vortex matter-wave 
interferometer ($lab$) with topological charge $\ell=\pm 2$ in different 
moving reference frames:  
a) $lab$ is rotated only with angular velocity $\vec \Omega_{\oplus}$. For 
positive $\ell$ atomic cloud rotates  
clockwise and it moves up. For negative $\ell$ atomic cloud rotates 
counter-clockwise and it moves down.
b) $lab$ is translated up without rotation 
with linear velocity $\vec V$. For positive $\ell$ atomic cloud rotates 
clockwise and it moves up. For negative $\ell$ atomic cloud rotates 
counter-clockwise and it moves down. Grey horizontal line indicates 
unmovable LG beam necklace. } 
\label{fig.2}
\end{figure} 

\end{document}